# Critical phenomena employed in hydrodynamic problems: A case study of Rayleigh-Bénard convection

Michel Assenheimer[†] and Victor Steinberg

*Department of Physics of Complex Systems*
*The Weizmann Institute of Science*
*76100 Rehovot*
*Israel*

(November 19, 2018)

By virtue of Rayleigh-Bénard convection, we illustrate the advantages of combining a hydrodynamic pattern forming instability with a thermodynamic critical point. This has already lead to many novel unexpected observations and is further shown to possess opportunities for the study of exciting fundamental problems in nonequilibrium systems.

## I. INTRODUCTION

Numerous systems, whose microscopic features differ substantially, can exhibit (quasi)periodic macroscopic spatial patterns when driven out of equilibrium [1]. Generally, they are associated with manifestations of nonlinear effects. The richness of structures in such spatially extended systems, is strikingly similar in systems of physical, hydrodynamic, chemical or biological origin [1].

Among the main goals in the study of pattern formation are understanding: (i) the universal mechanisms which govern the nonlinear spatio-temporal dynamics in various systems; (ii) their relation to the underlying symmetries; (iii) the various transitions from ordered to disordered states and their possible universality; and (iv) their characterization. Extensive studies of nonequilibrium pattern forming systems during the last decades revealed many qualitatively new dynamic phenomena. Many of these features are now being incorporated into theories in other fields such as superconductors and superfluids [2,3].

Although each experimental system has its own benefits, thermal convection of a thin horizontal fluid layer heated from below (i.e. Rayleigh-Bénard convection), offers exceptional experimental, theoretical and numerical advantages. Well-controlled experiments, with well-defined boundary conditions are easy to perform, and all relevant material properties of both the working fluids and boundaries are readily available. Not only are the underlying equations well-known theoretically, but the extensive and detailed nonlinear stability analysis, mainly performed by Busse and his collaborators [4], of an initially straight roll pattern, renders it the best documented pattern forming system (see inset).

Consequently, over the years Rayleigh-Bénard convection (RBC) proved to be an excellent model system in which to study pattern formation. There are, nevertheless, two serious drawbacks to RBC, common to most other systems as well. First, the experimental limitation to utilize large aspect ratio $\Gamma$ cells with ordinary working fluids, and second, the restriction to vary the fluid properties in a wide range. Of importance here is the Prandtl number $P$, which is the nondimensional ratio of the thermal and viscous time scales.

---

### Basics of Rayleigh-Bénard convection

In Rayleigh-Bénard convection (RBC) a thin horizontal fluid layer of thickness $d$ is heated from below. The nondimensional control parameters for this problem are the Rayleigh number $R$ and the Prandtl number $P$. The Rayleigh number $R \sim \Delta T\, d^3$ describes the external forcing, while $P$ represents the physical properties of the fluid. There exists a critical value for the stress across the layer, $R_c = 1708$ (and a corresponding critical temperature difference $\Delta T_c$), at which the bouyancy force exceeds the viscous and thermal dissipation. The resulting fluid motion, induced by the instability, is characterized by patterns with a typical size of the order of $d$ and wavenumber $k \sim d^{-1}$.

The detailed bifurcation sequence and pattern dynamics, however, depends on another two nondimensional quantities, namely the non-Boussinesq parameter $Q$ and the aspect ratio of the convection cell $\Gamma$. $Q$ determines the degree of deviation from the up-down symmetry of the RBC problem. $\Gamma$ is the ratio of a typical horizontal to vertical dimension of the cell. Here, the horizontal cell size is taken as the typical horizontal scale, although sometimes a size based on average interdefect distances is considered. In the latter case $\Gamma$ is dynamic, depending both on $R$ and $P$. When $\Gamma$ is sufficiently large, the spatial degrees of freedom define the pattern dynamics and the system is said to be *spatially extended*.

---

Rather early it was realized that RBC in a gas under high pressure allows for a significant reduction of the cell thickness by about a factor 2 to 3, yet without violating the Boussinesq approximation which assumes the fluid properties to be constant, except for the density variation in the driving bouyancy term [5]. Hence, for a fixed cell diameter larger aspect ratios can be obtained. Today's state of the art experiments use gases to attain $\Gamma$ close to about 100. In gas convection, however, $P \approx 1$.

Both disadvantages can be overcome easily by using a fluid near its liquid-vapor critical point (see inset). The strong and moreover particular temperature dependence of many fluid properties near the critical density, permits to scan $P$ continuously from unity to practically infinity. Moreover, the critical temperature difference for the onset of convection, $\Delta T_c$, asymptotically approaches zero when the thermodynamic critical point is neared [6]. Hence, extremely thin cells become feasible and consequently exceptionally large $\Gamma$. Thus, with a single fluid in large aspect ratio cells, it becomes possible to study pattern formation in a substantial fraction of $(R, P)$ space. Furthermore, investigations of pattern dynamics along a path of constant $R$ and variable $P$ now become feasible [7].

---

### *Thermodynamic critical phenomena*

The liquid-gas critical point $(T_c, p_c, \rho_c)$ (critical temperature, pressure and density respectively) is a particular location in thermodynamic phase space at which most thermodynamic and kinetic fluid properties exhibit universal anomalies. These singular contributions can be presented in the proximity of the critical point in the form of power laws of $\tau = (T - T_c)/T_c$ (at $\rho = \rho_c$) with known fluid independent universal exponents.

The characteristic distance (i.e. the correlation length $\xi = \xi_0 \tau^{-\nu}$) over which local density fluctuations are felt, can, sufficiently close to the critical point, even reach macroscopic sizes ($\xi_0$ is of the order of the intermolecular distance and $\nu \approx 0.63$). For example, at $\tau = 10^{-6}$, $\xi \approx 1$ $\mu$m, which can be comparable with $d$. Precisely this appearance of a new length scale is responsible for the universal, fluid independent critical behavior.

---

Although hydrodynamics near a critical point has been discussed in detail [6], pattern formation in this regime has received only limited attention. The strong variability of the fluid properties near $T_c$ was first exploited at low temperatures in convecting helium gas to vary $P$ up to 4 with $\Gamma = 57$ [8]. The first visualized pattern formation experiments, however, were performed by S. Fauve et al. and concern the interaction between a parametric instability of a liquid-vapor interface and a liquid-vapor phase transition near $T_c$ [9]. Very recently a new type of boundary layer instability in the close vicinity of $T_c$ was also observed [10].

We extensively exploited the advantages offered by a gas near $(\rho_c, T_c)$, by conducting RBC experiments in pressurized $SF_6$ near the gas-liquid critical point. Several fundamental problems can sucessfully be tackled in this system: $(i)$ pattern formation and selection in extremely large systems in a very wide range of parameter space and as a function of $P$ in particular; $(ii)$ effects of the strong coupling of thermodynamic fluctuations with the order parameter responsible for the bifurcation to the patterned state; and $(iii)$ behavior of a strongly fluctuating hydrodynamic system when the hydrodynamic size is on the order of the fluctuation size. In this brief review we mostly present results on the first issue.

## II. SPATIO-TEMPORAL PATTERNS IN EXTENDED RBC

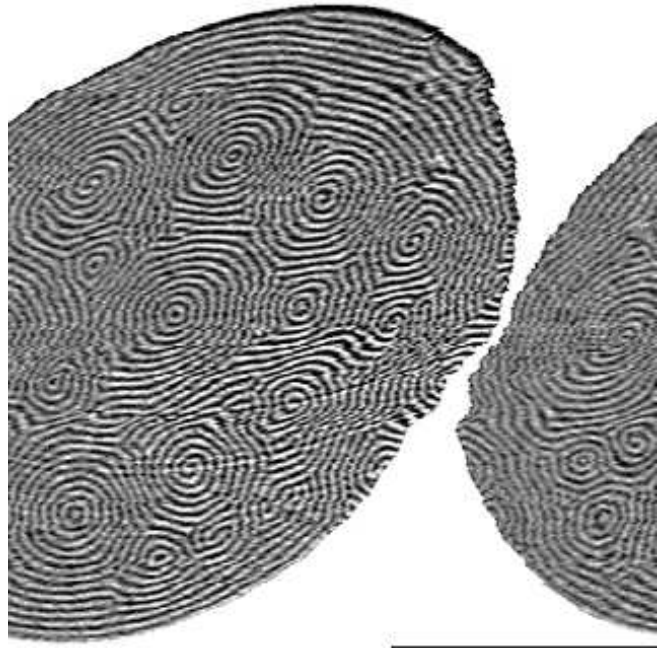

FIG. 1. *A typical spatio-temporal chaotic RBC state at $R = 1.95 R_c$ and $P = 6.6$ in a large aspect ratio convection cell with radial aspect ratio $\Gamma = 80$ and $d = 380 \mu m$. The convecting fluid is $SF_6$ at $T - T_c = 4.992$ K ($T_c = 318.73$ K) and density $\rho = 1.015 \rho_c = 748$ $kg/m^3$. The temperature difference across this thin layer is only $\Delta T = 17.4$ mK.*

One of the most intriguing recent discoveries in *natural* pattern formation (i.e. without externally imposed constraints on the degrees of freedom) is the recent observation of disordered spiral and target patterns [7,11,12] in large aspect ratio RBC, where previously only rolls were known to be stable [4]. Subsequently, these novel states were accurately reproduced both by numerical simulation of the generalized Swift-Hohenberg model [13], as well as by the integration of the full thermally driven Navier-Stokes equations in the Boussinesq approximation [14]. Although, it has since been established that these extended patterns are intrinsic to RBC, there is still little understanding of their dynamic behavior or the reasons why this state develops.

Our experimental studies have convincingly demonstrated the advantages of using a gas near the gas-liquid critical point for investigating this class of problems, particularly due to its uniqueness in reaching extremely large aspect ratios. For example, we are presently working with cells of 19 $\mu$m thickness which yield an aspect ratio close to 1000. On the other hand, the Prandtl number $P$ can be scanned over an extremely wide range, as can the parameter describing the non-Boussinesq behavior $Q$ [1,4]. These features have already lead to the observation of numerous unexpected phenomena including: novel extended patterns of many spirals and targets (Fig. 1); new mechanisms and scenarios for the evolution to disorder (Fig. 2) [15]; spiral-target transitions [12]; dislocation-core instabilities [7]; a breakup-reconnection mechanism responsible for the creation of the extended spirals and targets (Fig. 3) [7,12,15,16]; and the coexistence of up- and downflow hexagons (Fig. 4) [17].

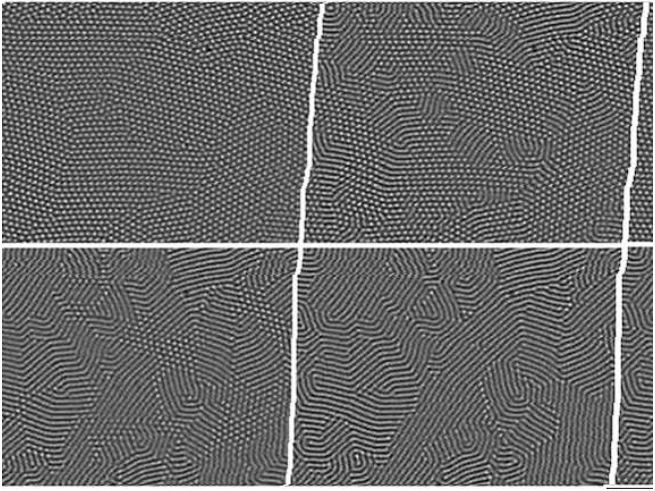

FIG. 2. *Transient evolution from a hexagonal state at $R \approx 1.5 R_c$ to a globally labyrinthine state at a slightly higher value of R. The labyrinthine state persists for at least one horizontal diffusion time (typical time for heat diffusion over the horizontal cell dimension). The experiment is performed at $P \sim O(100)$ very close to the critical point ($\tau \sim O(10^{-5})$) in a convection cell with $d = 19 \mu m$ and $\Gamma \approx 200$.*

These experiments raised a number of fundamental questions. For example: (*i*) What is the condition for the appearance of states containing many targets or many spirals? (*ii*) What is the mechanism of the spiral-target transition? (*iii*) What different classes of spatio-temporal behavior and spatio-temporal disordered states exist in nonequilibrium systems? (*iv*) What is the region of stability of the newly discovered states, and what is their location relatively to the well known stability region for straight roll patterns given by the so-called Busse stability balloon [1,4])? (*v*) What is the role of long range forces (which in RBC are created by horizontal pressure gradients) in pattern dynamics at high $P$ [18]? And finally, (*vi*) is it possible to experimentally observe an effect predicted long ago: the nonlinear interaction of thermodynamic fluctuations with the hydrodynamic order parameter leading to a weak first-order jump [19,20]?

Below we present three new types of patterns recently observed in RBC.

## III. TARGET-SPIRAL CHAOTIC STATES AT P > 1

A considerable step forward in the understanding of complex pattern dynamics in RBC was the appreciation of the crucial role played by the large scale mean drift flow [18], a nonlocal phenomenon. This mean flow is driven by pattern curvature which in turn advects and distorts the pattern itself. This effect is particularly relevant for low $P$ fluids [5,18]. While the relative importance of mean flow effects diminishes with increasing $P$, its weakness is compensated for by its nonlocal action over increasingly larger distances. Consequently, pattern dynamics, stability and selection can be altered considerably by mean flow, even at $P > 1$, in sufficiently large aspect ratio cells, a fact previously overlooked [1].

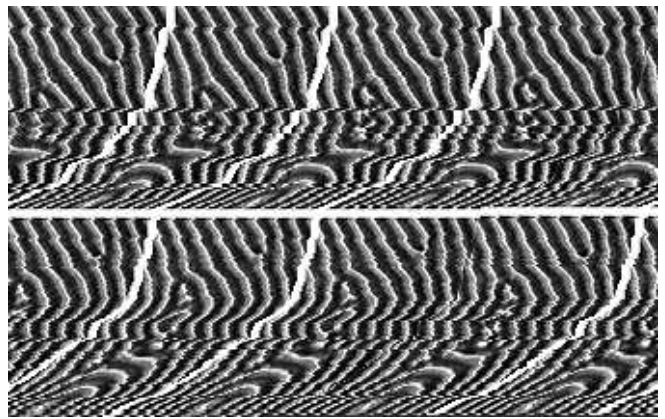

FIG. 3. *A typical breakup of a short enclosed roll initiated by wavenumber adaption of the background pattern, leading to the formation of a target. Time progresses from left to right and top to bottom in units of the vertical diffusion time.*

Thus the main ingredient which distinguishes various recent experiments from previous ones and which leads to spiral- and target-defect chaos is the large aspect ratio. Comparison of the experimental observations with simulations of the Swift-Hohenberg equation coupled with the mean drift equation at $P \approx 1$ shows that the large scale flow is responsible for spiral-defect chaos, and that up-down symmetry breaking (non-Boussinesq) effects are irrelevant [13,14].

So far the only theoretical attempt to clarify the occurence and characteristics of extended patterns, introduced the notion of *invasive defects* [21]. In this theory, wavenumber frustration is the dominant concept for understanding the dynamics. The defects, driven by the difference in wavenumber between that selected by a focal defect and some background wavenumber, expand to form spirals and targets and eventually take over the entire system. Cross and Tu also mention a novel core instability of axisymmetric targets to nonaxisymmetric perturbations which may induce a transition from a target morphology to a spiral. The latter observation can be related to the experimentally observed transition from spirals to targets as a function of $P$. It is actually quite remarkable that the generalized Swift-Hohenberg model so well reproduces the experimentally observed patterns as a function of the coupling to the mean flow (which is related to $P$). Similar results were obtained in numerical simulations of the full Navier-Stokes equations. [14]

Particularly because an established theoretical framework is lacking, the first obvious experimental step is to map out the stability region of the novel extended patterns in $(R, P, k)$ space. An extensive study of the detailed mechanism(s) responsible for the creation of these patterns should then follow. We have already investigated both issues [15]. Currently, we are verifying whether the background wavenumber suggested in the Cross and Tu theory is related to that relevant in dislocation dynamics.

## IV. COEXISTING UP- AND DOWNFLOW HEXAGONS IN RBC

Another unexpected observation occurs at the upper stability limit of extended patterns. Still in the region in which straight rolls were considered to be stable, hexagonal patterns appear via a core instability of spirals and targets, even for negligible values of $Q$. Although, the appearance of hexagons is usually related to an external up-down symmetry breaking (e.g. temperature dependent fluid properties), here they occur definitely in the Boussinesq condition. Their most striking feature, however, is the fact that both up- and downflow hexagons coexist simultaneously. Moreover, the hexagons introduce a new wavelength which substantially differs from the roll wavelength (see Fig. 4).

Recently Dewel et al. [22] suggested a model in which a $k = 0$ mode (suggested to be of thermodynamic origin) couples to a basic bifurcation to rolls and produces the observed patterns. We suggest that the zero mode might be of hydrodynamic origin, i.e. due to mean flow caused by curved rolls. The appearance of hexagons is in this case related to a self-induced symmetry breaking caused by the coupling of the large scale flow with the underlying instability. The observed ratio of roll to hexagon wavenumber, which agrees remarkably well with numerical simulations based on the model, strongly supports our suggestion. A recent numerical stability analysis of hexagonal patterns show that far from convection onset both types of hexagons are observable and stable at $P > 1$ with a wavelength larger than that of the coexisting rolls [23].

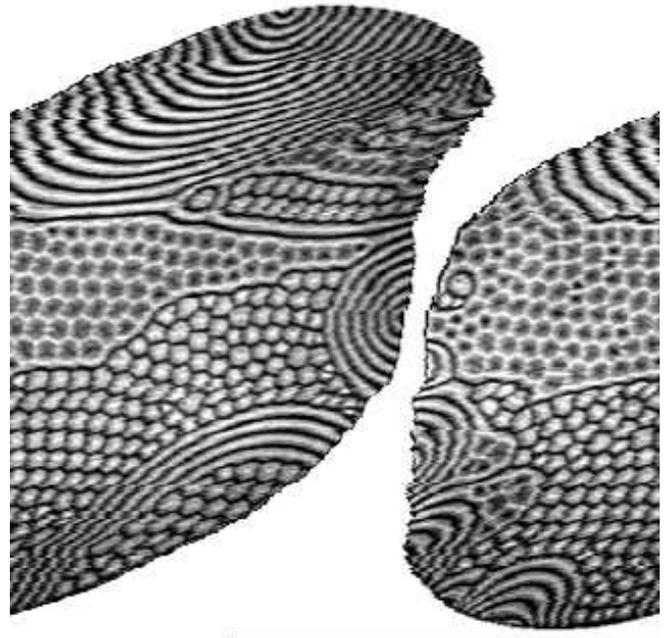

FIG. 4. *Typical coexisting up- and downflow hexagons together with rolls. Note that the hexagon wavelength is about 20% larger than the roll wavelength. The parameters are: $P = 4.5$, $R = 4.5R_c$, $\Gamma = 80$ and $d = 380 \mu m$.*

The observation of these hexagons seems to indicate that large scale flow is important not only for the transition to spatio-temporal complexity but also for the pattern and wavenumber selection problem itself.

## V. A NEW ROUTE FOR THE EVOLUTION OF DISORDER

A further interesting observation in very thin RBC cells (with $d \approx 20$ $\mu$m) is a new type of and the route to a globally isotropic disordered roll state which may be termed *labyrinthine* (see Fig. 2) [15]. This scenario differs from the usually observed hexagon to roll transition [1] presumably due to the non-negligible mean flow at large $R$ at which hexagons persist due to the large value of $Q$. Such a pattern reveals a remarkable structural similarity to states found in various *equilibrium* systems with competing interactions (i.e. amphiphilic monolayers, Type-I superconductors in the intermediate state, ferromagnetic garnet films, thin layers of ferrofluids in Hele-Shaw geometries, etc.), and in chemical reaction-diffusion systems [24].

The appearance of labyrinthine patterns requires the presence of nonlocal interactions. In the case of hydrodynamic patterns these can be associated with mean drift flows. Thus the observation of labyrinths at these large values of $P$ is another indication of the likely significance of large scale flow, contrary to the importance generally attributed to it for high $P$ fluids.

One of the mechanisms for the creation of labyrinths in magnetic garnet films is a transverse instability of ordered stripes. This instability eventually leads to the formation of a globally isotropic and disordered state via an unbinding of disclination dipoles [25]. We have not yet studied the generation of labyrinthine patterns in RBC, but a similar scenario can be expected. Indeed, the transverse instability observed in magnetic garnet film stripes [25] and the well-known zig-zag instability in RBC [1], are strikingly similar. In fact, the zig-zag instability in RBC is a mechanism by which a locally straight roll pattern of a wavelength larger than its optimal value, adapts its wavelength by alternate lateral bendings, hence the name: zig-zag. Globally, rolls will remain straight, however, with *zig* and *zag* excursions.

## VI. CONCLUDING REMARKS AND PROSPECTS

In this brief review we have tried to demonstrate the versatility and advantages of RBC near the critical point to study pattern formation in extremely large extended systems. The main result on the question of pattern and wavelength selection is the crucial - previously neglected - role played by the large scale mean flow and the transition to novel extended spatio-temporal chaotic states at $P > 1$. It should be emphasized that besides the widely accepted parameter $\epsilon/P^2$ ($\epsilon = (R - R_c)/R_c$) also $\Gamma^{-1}$ (here $\Gamma$ refers to the dynamic aspect ratio) plays a key role in mean flow strength. Although pointed out a long time ago, it was only appreciated recently [26]. The three distinct novel patterns, all ubiquitous in nature: extended targets and spirals, coexisting up- and down-flow hexagons and labyrinthine patterns all support this claim. In this respect, a further issue, namely the coarsening dynamics of the pattern upon a sudden jump from $R < R_c$ to $R > R_c$ can be used to probe the relevance of large scale mean flow, as has recently been suggested by Cross and Meiron [27].

Linear coupling between thermally induced fluctuations and a hydrodynamic bifurcation has recently been investigated [28]. In the vicinity of the critical point, the coupling is expected to become significantly nonlinear and might lead to a weak first-order jump in the bifurcation, even in the Boussinesq case. Preliminary experimental results indicate that this is indeed the case.

Close to the critical point, the increased thermodynamic correlation length can become comparable with $d$. Then, the hydrodynamic description of RBC might fail and become questionable altogether. This unique situation in which $\xi \sim O(d)$ will provide information about the dynamics of strongly fluctuating hydrodynamic flow.

---

*An introductory textbook on pattern formation in non-equilibrium systems is*

P. Manneville, *Dissipative Structures and Weak Turbulence*, Academic Press, Inc. (1990).

---

[†] Present address: Laboratoire de Physique Statistique, Ecole Normale Supérieure, 24 rue Lhomond, 75231 Paris Cedex 05, France. Email: fnassen@physique.ens.fr.
[1] For an advanced review see M. C. Cross and P. C. Hohenberg, Rev. Mod. Phys., **65**, 851 (1993).
[2] I. Aranson and V. Steinberg, to be published.
[3] I. Aranson, B. Shapiro, and V. Vinakur, Phys. Rev. Lett., **76**, 142 (1996).
[4] F. H. Busse, Rep. Prog. Phys., **41**, 1929 (1978); and references therein.
[5] V. Croquette, Contemp. Phys., **30**, 113 (1989).
[6] M. Gitterman and V. Steinberg, High Temperature (USSR), **8**, 754 (1970); for a review, see M. Gitterman, Rev. Mod. Phys., **50**, 85 (1978).
[7] M. Assenheimer and V. Steinberg, Phys. Rev. Lett., **70**, 3888 (1993).
[8] G. Ahlers and R. P. Behringer, Phys. Rev. Lett., **40**, 712 (1978).
[9] S. Fauve, K. Kumar, C. Laroche, D. Beysens, and Y. Garrabos, Phys. Rev. Lett., **68**, 3160 (1992).
[10] G. Buschhorn, U. Kilgus, W. Mark, H. Posselt, and J. Rubach, Europhys. Lett., to appear.
[11] S. W. Morris, E. Bodenschatz, D. S. Cannell, and G. Ahlers, Phys. Rev. Lett. **71**, 2026 (1993).
[12] M. Assenheimer and V. Steinberg, Nature, **367**, 345 (1994).
[13] H. Xi, J. D. Gunton, and J. Viñals, Phys. Rev. E, **47**, R2987 (1993); Phys. Rev. Lett., **71**, 2030 (1993); M. Beste-

horn, M. Fantz, R. Friedrich, and H. Haken, Phys. Lett. A, **174**, 48 (1993).
- [14] W. Decker, W. Pesch, and A. Weber, Phys. Rev. Lett., **73**, 648 (1994); private communication.
- [15] M. Assenheimer and V. Steinberg, in preparation.
- [16] M. Assenheimer, Ph. D. thesis, Weizmann Institute of Science (1994).
- [17] M. Assenheimer and V. Steinberg, Phys. Rev. Lett., to appear.
- [18] E. D. Siggia and A. Zippelius, Phys. Rev. Lett., **47**, 835 (1981).
- [19] S. A. Brazovskii, Sov. Phys. JETP, **41**, 85 (1975).
- [20] J. B. Swift and P. C. Hohenberg, Phys. Rev. A, **15**, 319 (1977); P. C. Hohenberg and J. B. Swift, Phys. Rev. E, **52**, 1828 (1995).
- [21] M. C. Cross and Y. Tu, Phys. Rev. Lett., **75**, 834 (1995).
- [22] G. Dewel, S. Métens, M'F. Hilali, P. Borckmans, and C. B. Price, Phys. Rev. Lett., **74**, 4647 (1995).
- [23] R. M. Clever and F. H. Busse, preprint (1995).
- [24] M. Seul and D. Andelman, Science, **267**, 476 (1995) and references therein.
- [25] M. Seul and R. Wolfe, Phys. Rev. Lett., **68**, 2460 (1992).
- [26] A. C. Newell, T. Passot, and M. Souli, Phys. Rev. Lett., **64**, 2378 (1990).
- [27] M. C. Cross and D. I. Meiron, Phys. Rev. Lett., **75**, 2152 (1995).
- [28] M. Wu, G. Ahlers, and D. S. Cannell, Phys. Rev. Lett., **75**, 1743 (1995).